\documentclass[12pt]{article}

\catcode`\@=11

\global\arraycolsep=2pt
\usepackage{amsbsy,amssymb,latexsym,amsfonts,amsmath}
\usepackage{graphicx,color}

\begin{document}
\begin{flushright}
\parbox{4.2cm}
{CALT 68-2934}
\end{flushright}

\vspace*{0.7cm}

\begin{center}
{ \Large $(0,2)$ Chiral Liouville Field Theory}
\vspace*{1.5cm}\\
{Yu Nakayama}
\end{center}
\vspace*{1.0cm}
\begin{center}
{\it California Institute of Technology,  \\ 
452-48, Pasadena, California 91125, USA}
\vspace{3.8cm}
\end{center}

\begin{abstract}
As an existence proof of the $(0,2)$ heterotic supercurrent supermultiplets in $(1+1)$ dimensional quantum field theories which are consistent with the warped superconformal algebra, we construct the $(0,2)$ chiral Liouville field theories.  The two distinct possibilities of the heterotic supercurrent supermultiplets are both realized.
\end{abstract}

\thispagestyle{empty} 

\setcounter{page}{0}

\newpage

\section{Introduction}
Scale invariance is ubiquitous in nature, and there has been some renewed interest in the subject without assuming Lorentz invariance. The assumption of Lorentz invariance naturally makes scale invariant quantum field theories conformal invariant. This is proved in $1+1$ dimension, and there is good evidence to believe the similar symmetry enhancement in higher dimensions (see e.g. \cite{Nakayama:2013is} for a review). The situation without Lorentz invariance is more subtle and interesting. It allows more exotic possibilities. 

In $1+1$ dimension, Hofman and Strominger argued \cite{Hofman:2011zj} that under certain technical assumptions, chiral scale invariance must be enhanced to chiral conformal invariance.
 We do not know whether their assumptions are plausible or not without any concrete realization. While there exist some holographic examples based on Kerr/CFT correspondence \cite{Guica:2008mu} and the generic holographic argument \cite{Nakayama:2011fe}, the concrete field theory realization has been unavailable. From the holographic construction, the algebra is also known as ``warped conformal algebra". Some field theory aspects have been investigated in \cite{Detournay:2012pc} with its focus on the modular property.

Although it may not satisfy all of the technical assumptions, in the latest paper \cite{Compere:2013aya}, the chiral Liouville field theory that is equipped with the warped conformal algebra was proposed as a particular gauge fixing of the conventional (quantum) gravity in $1+1$ dimension. It may have some interesting applications in the holographic dual in a similar way that the ordinary Liouville field theory can be regarded as the effective  boundary theory of the $\mathrm{AdS}_3$ space-time \cite{Compere:2013bya}.

In this paper, we construct the $(0,2)$ heterotic supersymmetric generalization of the chiral Liouville field theory. The $(0,2)$ heterotic supersymmetry compatible with the warped conformal algebra was classified in \cite{Nakayama:2013coa}, in which we have showed the two distinct possibilities aside from the trivial superconformal algebra with the Lorentz invariance. The goal of this paper is to give an existing proof of the $(0,2)$ heterotic supercurrent supermultiplets by constructing the $(0,2)$ supersymmetric generalization of the  chiral Liouville field theory. 
In section 2, we review the possible $(0,2)$ heterotic supercurrent supermultiplets compatible with the warped conformal algebra. The trivial case is the Lorentz invariant situation, and the construction of the Lorentz invariant $(0,2)$ Liouville field  theory can be found in \cite{Nakayama:2008fi}. In addition, there are two non-trivial possibilities.
In section 3 and 4, we construct the $(0,2)$ chiral Liouville field theory with and without the Liouville (super)potential respectively.  The two distinct possibilities of the heterotic supercurrent supermultiplets are both realized.

\section{$(0,2)$ heterotic supercurrent supermultiplets with warped superconformal algebra}

In \cite{Nakayama:2013coa}, we have classified the possible $(0,2)$ heterotic supercurrent supermultiplets, which are consistent with the warped conformal algebra \cite{Hofman:2011zj}. Aside from the Lorentz invariant superconformal field theories, we have two distinct possibilities. 

The first case is the situation in which the supersymmetry is realized in the chiral conformal sector. With the notation in \cite{Nakayama:2013coa}, the supercurrent supermultiplet must satisfy
\begin{align}
\partial_{--} \mathcal{S}_{++} &= 0 \cr
\partial_{--} Y_{-} & = 0 \cr
X_{-} & = \mathcal{T}_{----} = 0 \ , \label{righta}
\end{align}
where $\mathcal{S}_{++}$ and $\mathcal{T}_{----}$ are real superfields and $X_-$ and $Y_-$ are fermionic chiral superfields.
In components, we have the current algebra
\begin{align}
\partial_{--} j_{++} &= 0 \cr
\partial_{--} S_{+++} & = 0 \cr
\partial_{--} T_{++++}& = 0 \ , 
\end{align}
which generates the right $\mathcal{N}=2$ superconformal algebra (with super Virasoro extension) from $\mathcal{S}_{++}$ including energy-momentum tensor $T_{+++}$, supercurrent $S_{+++}$ and the R-current $j_{++}$, and the right $\mathcal{N}=2$ Kac-Moody current from $Y_{-}$,
which generates the {\it left} translation (together with additional bosonic and fermionic current algebra extension) from $P_{--} = \int dx^{++} T_{++--}$ with $\partial_{--} T_{++--} = 0$.

The other possibility is the situation in which the supersymmetry is realized in the  opposite sector to the chiral conformal algebra. 
 The supercurrent supermultiplet must satisfy
\begin{align}
D_+ X_- - \bar{D}_+ \bar{X}_- &= 0 \cr
\bar{D}_{+} \mathcal{T}_{----} &= 0 \cr
Y_{-} & = \mathcal{S}_{++} = 0 \ . \label{lefta}
\end{align}
In components, we have the current algebra
\begin{align}
\partial_{++} T_{----} &= 0 
\end{align}
which generates the left conformal algebra (with Virasoro extension) from $\mathcal{T}_{----}$. 
Note that the energy-momentum tensor $T_{----}$ is singlet under the $(0,2)$ supersymmetry. We also have the left Kac-Moody current 
which generates the {\it right} supersymmetry algebra (together with bosonic and fermionic current algebra extension) from $X_-$. The right translation is given by  $P_{++} = \int dx^{--} T_{--++}$ with $\partial_{++} T_{--++} = 0$.

\section{$(0,2)$ chiral Liouville field theory without potential}
In \cite{Compere:2013aya}, a novel gauge fixing of $(1+1)$ dimensional gravity was proposed, which leads to the so-called chiral  Liouville field theory.\footnote{We present the left realization of the warped conformal algebra here. It is trivial to construct the right realization by exchanging $x^{++}$ and $x^{--}$, which will be used in section 3.3 with the supersymmetry.} The chiral Liouville field theory has a non-covariant and non-local ``action" given by
\begin{align}
S_0 &= \int d^2x \left(\partial_{++} \rho \partial_{--} \rho - \frac{\Lambda}{8}e^{2\rho} + h(\partial_{--} \rho)^2 \right.) \cr
& \left. + [\partial_{--} h \partial_{--} \rho + \frac{1}{2} \int d^2 x' \partial_{--}^2 h(x) G(x,{x'}) \partial^2_{--} h(x')] \right) \ , \label{cla}
\end{align}
where $G(x,x') = \frac{1}{2\pi} \log|x-x'|$ is $(1+1)$ dimensional Green's function. 

Strictly speaking, the chiral Liouville field theory does not have the action principle and the above ``action" is only mnemonic to derive the equations of motion by varying fields $h$ and $\rho$. After the variation, we impose the constraint $\partial_{--} h = 0$, which makes the last line in \eqref{cla} vanish. Furthermore, to make the left translation zero-mode integral better defined, we will consider the fixed zero mode action \cite{Compere:2013aya} by adding the (non-covariant) term
\begin{align}
S_L = S_0 -\Delta\int d^2x h \ .
\end{align}
The complete set of equations of motion and the constraint is
\begin{align}
\partial_{++} \partial_{--} \rho &= -\frac{\Lambda}{8}e^{2\rho} - h \partial_{--}^2 \rho \cr
\partial_{--}\rho^2 - \partial^2_{--} \rho - \Delta &= 0 \cr
\partial_{--} h &= 0 \ .
\end{align}

The theory possesses the ``traceless" left energy-momentum tensor
\begin{align}
\partial_{--} T_{++++} &= 0 \cr
T_{++++} & = (\partial_{++}\rho)^2 - \partial^2_{++} \rho + 2h(\partial_{++} \rho \partial_{--} \rho - \partial_{++} \partial_{--} \rho) -\partial_{++} h \partial_{--} \rho \  
\end{align}
from which one can construct the left momentum $P_{++} = \int dx^{++} T_{++++}$. On the other hand, the right translation is generated by the Kac-Moody current 
\begin{align}
\partial_{--} T_{++--} & = 0 \cr
T_{++--} & = 2 \Delta h 
\end{align}
from which one can construct the right momentum $P_{--} = \int dx^{++} T_{++--}$.\footnote{We see that when $\Delta =0$, the right momentum becomes trivial, and the Dirac bracket becomes singular \cite{Compere:2013aya}.}

In the following, we will construct the $(0,2)$ supersymmetric version of the chiral Liouville field theory. As recalled in section 2, there are two distinct supersymmetric realizations of the $(0,2)$ supersymmetry algebra consistent with the warped conformal algebra. In section 3.2 we construct the left realization of the warped conformal symmetry, and in section 3.3 we construct the right realization of the warped conformal symmetry. Throughout this section, we persue the minimal possibility based on one chiral multiplet $\Phi$ (a generalization of $\rho$) and one chiral constrained multiplet $\Psi$ (a generalization of $h$). The minimal construction does not allow the supersymmetric analogue of the Liouville potential term. The introduction of the Liouville potential term will be discussed in section 4 with the addition of a Fermi-multiplet.

\subsection{Left realization}
The superspace action is given by\footnote{Our convention is $\int d^2\theta (\theta^+ \bar{\theta}^+) =\int d\theta^+ (\theta^+) = \int d \bar{\theta}^+ (\bar{\theta}^+) = 1$.}
\begin{align}
S &= \int d^2x  d^2\theta \left(-i \Phi \partial_{--} \bar{\Phi} \right) \cr
 &+ \int d^2x d\theta^+ (\Psi [(\partial_{--}\Phi)^2 - (\partial_{--}^2 \Phi) + \Delta]) \cr
&+ \int d^2x d\bar{\theta}^+ (\bar{\Psi} [(\partial_{--}\bar{\Phi})^2 - (\partial_{--}^2 \bar{\Phi}) + \bar{\Delta}]) \ . \label{left}
\end{align}
With the component expansion
\begin{align}
\Phi &= \phi + i \theta^+ \lambda + \theta^+ \bar{\theta}^+ (\frac{i}{2}\partial_{++} \phi) \cr
\bar{\Phi} &= \phi^* + i \bar{\theta}^+ \bar{\lambda} + \theta^+ \bar{\theta}^+ (-\frac{i}{2}\partial_{++} \phi) \cr
\Psi &= \psi + \theta^+ h - \theta^+ \bar{\theta}^+ (\frac{i}{2}\partial_{++} \psi) \cr
\bar{\Psi} &= \bar{\psi} + \bar{\theta}^+ h^* + \theta^+ \bar{\theta}^+ (\frac{i}{2}\partial_{++} \bar{\psi}) 
\end{align}
we have 
\begin{align}
S &= \int d^2x \left( \partial_{++} \phi \partial_{--} \phi^* - i\lambda \partial_{--} \bar{\lambda}  \right. \cr
& + h[(\partial_{--}\phi)^2 - (\partial^2_{--} \phi) + \Delta] + h^* [(\partial_{--} \phi^*)^2 - (\partial^2_{--} \phi^*) + \Delta^*] \cr
& \left. + \psi [-2i (\partial_{--} \lambda) (\partial_{--}\phi) + i \partial_{--}^2 \lambda] + \bar{\psi} [-2i(\partial_{--}\bar{\lambda}) (\partial \phi^*) + i\partial_{--}^2 \bar{\lambda}] \right) \ .
\end{align}

As in the bosonic case, we should regard the action as mnemonic to derive the equations of motion by varying $\Phi$ and $\Psi$. Only after the variation, we impose the chirality constraint $\partial_{--} \Psi = 0$. To focus on the fixed energy sector, a complex number $\Delta$ is introduced.
Furthermore, we have suppressed a possible non-local term, which does not alter the equations of motion with the constraint. 
The set of equations of motion and the constraint is summarized as
\begin{align}
(\partial_{--} \Phi)^2 - (\partial_{--}^2 \Phi) + \Delta &= 0 \cr
i\partial_{--}\bar{D}_+ \bar{\Phi} + 2(\partial^2_{--} \Phi) \Psi  &= 0 \cr
\partial_{--} \Psi & = 0 \ .
\end{align}

The left supercurrent supermultiplet is given by the sum of the two separately conserved ones:
\begin{align}
\mathcal{S}_{++} = \mathcal{S}^{(0)}_{++} + \alpha \mathcal{S}^{\mathrm{imp}}_{++} \ . \label{leftsuper}
\end{align}
Here
\begin{align}
\mathcal{S}^{(0)}_{++} &= D_+ \Phi \bar{D}_+ \bar{\Phi} + 2i \Psi (D_+\Phi) (\partial_{--} \Phi) +2i \bar{\Psi} (\bar{D}_+\bar{\Phi}) (\partial_{--} \bar{\Phi}) \cr 
& -i \Psi (\partial_{--} D_+ \Phi) - i \bar{\Psi} (\partial_{--} \bar{D}_+ \bar{\Phi}) \ .
\end{align}
The improvement ambiguity, which does not change the space-time translation charges, is given by 
\begin{align}
\mathcal{S}_{++}^{\mathrm{imp}}  &= i \partial_{++}\Phi - i \partial_{++} \bar{\Phi} +2i \Psi [\partial_{--} D_+ \Phi] + 2i \bar{\Psi}[\partial_{--} \bar{D}_+ \bar{\Phi}] \cr 
& + 2i (D_+ \Psi) \partial_{--}\Phi - 2i (\bar{D}_+ \bar{\Psi}) \partial_{--} \bar{\Phi} 
 .
\end{align}
Note that the top component of $\mathcal{S}_{++}$ is the R-current, and the above improvement changes the R-symmetry.

The additional Kac-Moody current is realized by
\begin{align}
Y_{-} = \Delta \Psi \ ,
\end{align}
which satisfies $\partial_{--} Y_{-} = 0$ from the constraint. 
With the usage of the equations of motion and the constraint, we can show that these satisfies the supercurrent conservation \eqref{righta}. 

\subsection{Right realization}
The superspace action is given by
\begin{align}
S &= \int d^2x d^2\theta \left(-i \Phi \partial_{--} \bar{\Phi} \right) \cr
 &+ \int d^2x d\theta^+ (\Psi [(\partial_{++}\Phi)^2 - (\partial_{++}^2 \Phi) + \Delta]) \cr
&+ \int d^2x d\bar{\theta}^+ (\bar{\Psi} [(\partial_{++}\bar{\Phi})^2 - (\partial_{++}^2 \bar{\Phi}) + \bar{\Delta}]) \ . \label{right}
\end{align}
With the component expansion
\begin{align}
\Phi &= \phi + i \theta^+ \lambda + \theta^+ \bar{\theta}^+ (\frac{i}{2}\partial_{++} \phi) \cr
\bar{\Phi} &= \phi^* + i \bar{\theta}^+ \bar{\lambda} + \theta^+ \bar{\theta}^+ (-\frac{i}{2}\partial_{++} \phi) \cr
\Psi &= \psi + \theta^+ h - \theta^+ \bar{\theta}^+ (\frac{i}{2}\partial_{++} \psi) \cr
\bar{\Psi} &= \bar{\psi} + \bar{\theta}^+ h^* + \theta^+ \bar{\theta}^+ (\frac{i}{2}\partial_{++} \bar{\psi}) 
\end{align}
we have 
\begin{align}
S &= \int d^2x \left( \partial_{++} \phi \partial_{--} \phi^* - i\lambda \partial_{--} \bar{\lambda} \right. \cr
& + h[(\partial_{++}\phi)^2 - (\partial^2_{++} \phi) + \Delta] + h^* [(\partial_{++} \phi^*)^2 - (\partial^2_{++} \phi^*) + \Delta^*] \cr
& \left. + \psi [-2i (\partial_{++} \lambda) (\partial_{++}\phi) + i \partial_{++}^2 \lambda] + \bar{\psi} [-2i(\partial_{++}\bar{\lambda}) (\partial_{++} \phi^*) + i\partial_{++}^2 \bar{\lambda}] \right) \ .
\end{align}

Again, we should regard the action as mnemonic to derive the equations of motion by varying $\Phi$ and $\Psi$. Only after the variation, we impose the chirality constraint $\Delta D_{+} \Psi - \bar{\Delta} \bar{D}_+ \bar{\Psi}  = 0$.  To focus on the fixed energy sector, a complex number $\Delta$ is introduced. Furthermore, we have suppressed a possible non-local term, which does not alter the equations of motion with the constraint.
The set of equations of motion and the constraint is summarized as 
\begin{align}
(\partial_{++} \Phi)^2 - (\partial_{++}^2 \Phi) + \Delta &= 0 \cr
i\partial_{--}\bar{D}_+ \bar{\Phi} + 2(\partial^2_{++} \Phi) \Psi  &= 0 \cr
\Delta D_{+} \Psi - \bar{\Delta} \bar{D}_+ \bar{\Psi} & = 0 \ .
\end{align}
The last equation implies $\partial_{++} \Psi = 0$.

The right supercurrent supermultiplet is given by the sum of the two separately conserved ones:
\begin{align}
\mathcal{T}_{----} = \mathcal{T}^{(0)}_{----} + \tilde{\alpha} \mathcal{T}^{\mathrm{imp}}_{----} \ . \label{rightsuper}
\end{align}
Here
\begin{align}
\mathcal{T}^{(0)}_{----} = &(\partial_{--} \Phi) (\partial_{--} \bar{\Phi}) - 2 (\partial_{++} D_+\Phi)\Psi (\partial_{--}\Phi)  - 2 (\partial_{++} \bar{D}_+ \bar{\Phi}) \bar{\Psi} (\partial_{--} \bar{\Phi}) \cr
&-2(\partial_{++} \Phi) \Psi (\partial_{--} D_+ \Phi) + 2(\partial_{++}\bar{\Phi}) \bar{\Psi} (\partial_{--} \bar{D}_+ \bar{\Phi}) \cr
&+2 (\partial_{++}\Phi) (D_+ \Psi) (\partial_{--} \Phi) -2 (\partial_{++} \bar{\Phi} )(\bar{D}_+ \bar{\Psi}) (\partial_{--} \bar{\Phi}) \cr
&+ (\partial_{++} \partial_{--} D_+ \Phi)\Psi - (\partial_{++} \partial_{--} \bar{D}_+ \bar{\Phi}) \bar{\Psi} \cr
&-(\partial_{++} \partial_{--}\Phi) (D_+ \Psi) + (\partial_{++}\partial_{--} \bar{\Phi}) (\bar{D}_+ \bar{\Psi}) \ .
\end{align}
The improvement ambiguity, which does not change the space-time translation charges, is given by 
\begin{align}
\mathcal{T}_{----}^{\mathrm{imp}}  = & \partial^2_{--}\Phi + \partial^2_{--}\bar{\Phi}  -2 (\partial_{++}\partial_{--} D_+ \Phi) \Psi +2 (\partial_{++} \partial_{--} \bar{D}_+ \bar{\Phi} ) \bar{\Psi} \cr
&+ 2(\partial_{++}\partial_{--} \Phi) (D_+ \Psi) - 2(\partial_{++} \partial_{--} \bar{\Phi}) (\bar{D}_+\bar{\Psi}) \cr
&-2(\partial_{++} D_+ \Phi) (\partial_{--}\Psi) + 2(\partial_{++} \bar{D}_+ \bar{\Phi}) (\partial_{--} \bar{\Psi}) \cr
&+2 (\partial_{++} \Phi) (\partial_{--}D_+ \Psi) -2 (\partial_{++} \bar{\Phi} ) (\partial_{--} \bar{D}_+ \bar{\Psi}) \ . 
\end{align}

The additional Kac-Moody current is realized by
\begin{align}
X_- = \Delta \Psi \ ,
\end{align}
which satisfies the condition $D_+ X_- - \bar{D}_+ \bar{X}_- = 0$ from the constraint. 
With the usage of the equations of motion and the constraint, we can show that these satisfies the supercurrent conservation \eqref{lefta}.

\section{Introduction of Liouville potential}
One feature of the construction of the $(0,2)$ chiral Liouville field theory in the last section was there is one-parameter freedom to choose the energy-momentum tensor. Another feature was there is no Liouville potential. These are related because, as we will see in this section, the improvement ambiguity is fixed by the Liouville potential in (chiral) Liouville field theories.

However, we should note that it is impossible to write down supersymmetric Liouville potential term without introducing additional degrees of freedom. For this purpose, we introduce a Fermi-multiplet $\Gamma$, which is a fermionic chiral superfield (i.e. $\bar{D}_+ \Gamma = 0$), with the component expansion
\begin{align}
\Gamma = \gamma + \theta^+ F - \theta^+\bar{\theta}^+(\frac{i}{2}\partial_{++} \gamma) \ .
\end{align}
The kinetic term of the Fermi-multiplet is given by
\begin{align}
S_{\mathrm{kin}} = \int d^2x d^2 \theta \bar{\Gamma} \Gamma = \int d^2x \left( -\frac{i}{2} \bar{\gamma} (\partial_{++} \gamma) +\frac{i}{2}(\partial_{++}\bar{\gamma})\gamma - \bar{F}F \right) \ , \label{kine}
\end{align}\
where $F$ plays the role of the auxiliary field.

The $(0,2)$ Liouville potential can be introduced as
\begin{align}
S_{\mathrm{pot}} &= \mu \int d^2x d \theta^+ \Gamma e^{b \Phi} + \bar{\mu} \int d^2x d \bar{\theta}^+ \bar{\Gamma} e^{b\bar{\Phi}} \cr
& = \int d^2x \left( \mu Fe^{b\phi} + \bar{\mu} \bar{F} e^{b{\phi}^*} - ib\mu\gamma\lambda e^{b\phi} - ib \bar{\mu} \bar{\gamma} \bar{\lambda} e^{b \phi^*} \right) , \label{poten}
\end{align}
where $\mu$ is complex, but $b$ is a real parameter. We can add the kinetic term \eqref{kine} and the potential term \eqref{poten} both in the left realization  \eqref{left} and the right realization \eqref{right} of the $(0,2)$ warped conformal field theories constructed in the last section. The equations of motion can be obtained by varying $\Phi$, $\Psi$ and $\Gamma$, and then we impose the constraint on $\Psi$.

The warped conformal algebra now relates the real number $b$ with the improvement ambiguity $\alpha$ and $\tilde{\alpha}$ in the supercurrent supermultiplets. In the left realization, the necessity can be seen from the fact that $\gamma$ has the zero chiral scaling dimension, and $\psi$ has one half, so $e^{b\phi}$ should have additional one half of the chiral scaling dimension by adjusting the shift transformation under the chiral dilatation induced by the improved definition of the left energy-momentum tensor.\footnote{The same result is obtained from the R-symmetry consideration.} This determines $\alpha = -\frac{1}{b}$. 
Alternatively, in the right realization, $\gamma$ has one half of the chiral scaling dimension, and $\lambda$ has the zero chiral scaling dimension, so again $e^{b\phi}$ should have additional one half of the chiral scaling dimension by adjusting the shift transformation under the chiral dilatation induced by the improved definition of the right energy-momentum tensor. This fixes $\tilde{\alpha} = -\frac{1}{2b}$.

We can verify the (classical) sufficiency of the $(0,2)$ warped superconformal invariance by constructing the explicit $(0,2)$ supercurrent supermultiplets. They take the same form as \eqref{leftsuper} and \eqref{rightsuper} with the above specified values of $\alpha$ and $\tilde{\alpha}$. By using the modified equations of motion with the Liouville potential terms, one can see that it satisfies the conservation condition.

\section{Conclusion}
In this paper, we have constructed the $(0,2)$ chiral Liouville field  theories as an existence proof of the $(0,2)$ heterotic supercurrent supermultiplets in $(1+1)$ dimensional quantum field theories which are consistent with the warped superconformal algebra. The two distinct possibilities of the heterotic supercurrent supermultiplets are both realized. 

In both cases, the bosonic part can be regarded as the generalization of the chiral Liouville field theory. On the other hand, the fermionic part shows an interesting distinction. In the left realization, with neglecting the interaction, it is given by chiral fermions. On the other hand, in the right realization, with neglecting the interaction, it is essentially given by the quantum mechanical zero mode algebra of fermionic harmonic oscillator with zero frequency (see \cite{Nakayama:2012ed} for a related comment on the possibility of such a realization).

Our construction is classical, and it is extremely important to address the question of the quantum warped conformal invariance by studying the quantization of the chiral Liouville field theories. It is particularly of interest to compute the central charges for the $(0,2)$ warped superconformal algebra. In the bosonic case, it is expected that we would get the quantum corrections from the renormalization of the exponential Liouville operator much like in the conventional Liouville field theory (see e.g. \cite{Nakayama:2004vk} and reference therein). With the $(0,2)$ supersymmetry and the possible perturbative non-renormalization of the superpotential term,  we would expect that the Liouville potential term and the central charge is not renormalized here, but we may need a more rigorous argument to support the claim. In particular, there could exist non-perturbative effects that can be important in $(0,2)$ supersymmetric field theories. We would like to come back to these issues in the near future.

\section*{Acknowledgements}
This work is supported by Sherman Fairchild Senior Research Fellowship at California Institute of Technology  and DOE grant DE-FG02-92ER40701


\begin{thebibliography}{99}

\bibitem{Nakayama:2013is} 
  Y.~Nakayama,
  arXiv:1302.0884 [hep-th].






\bibitem{Hofman:2011zj} 
  D.~M.~Hofman and A.~Strominger,
  Phys.\ Rev.\ Lett.\  {\bf 107}, 161601 (2011)
  [arXiv:1107.2917 [hep-th]].





\bibitem{Guica:2008mu} 
  M.~Guica, T.~Hartman, W.~Song and A.~Strominger,
  Phys.\ Rev.\ D {\bf 80}, 124008 (2009)
  [arXiv:0809.4266 [hep-th]].


\bibitem{Nakayama:2011fe} 
  Y.~Nakayama,
  Phys.\ Rev.\ D {\bf 85}, 085032 (2012)
  [arXiv:1112.0635 [hep-th]].


\bibitem{Detournay:2012pc} 
  S.~Detournay, T.~Hartman and D.~M.~Hofman,
  Phys.\ Rev.\ D {\bf 86}, 124018 (2012)
  [arXiv:1210.0539 [hep-th]].


\bibitem{Compere:2013aya} 
  G.~Compere, W.~Song and A.~Strominger,
  arXiv:1303.2660 [hep-th].





\bibitem{Compere:2013bya} 
  G.~Compere, W.~Song and A.~Strominger,
  arXiv:1303.2662 [hep-th].

\bibitem{Nakayama:2013coa} 
  Y.~Nakayama,
  arXiv:1305.2937 [hep-th].


\bibitem{Nakayama:2008fi} 
  Y.~Nakayama,
  JHEP {\bf 0903}, 062 (2009)
  [arXiv:0810.4160 [hep-th]].

\bibitem{Nakayama:2012ed} 
  Y.~Nakayama,
  Phys.\ Rev.\ D {\bf 87}, 046005 (2013)
  [arXiv:1210.6439 [hep-th]].


\bibitem{Nakayama:2004vk} 
  Y.~Nakayama,
  Int.\ J.\ Mod.\ Phys.\ A {\bf 19}, 2771 (2004)
  [hep-th/0402009].

\end{thebibliography}
\end{document}